\documentclass{aa} 
\usepackage{graphics} 
\usepackage{amssymb}

\def\be{\begin{equation}}
\def\en{\end{equation}}

\def \ga {\mathrel{\vcenter
     {\offinterlineskip \hbox{$>$}\hbox{$\sim$}}}}
\input{psfig.sty}
\begin{document}

\title{Spherical collapse of supermassive stars: neutrino emission and
  gamma-ray bursts}

\author{Felix Linke \inst{1} 
\and Jos\'e A. Font \inst{1} 
\and Hans-Thomas Janka \inst{1} 
\and Ewald M\"uller \inst{1}
\and Philippos Papadopoulos \inst{2}}

\offprints{Jos\'e A. Font}
\mail{}

\institute{Max-Planck-Institut f\"ur Astrophysik,
Karl-Schwarzschild-Str. 1, 85741 Garching, Germany
\and
School of Computer Science and Mathematics, University of
Portsmouth, P01 2EG, Portsmouth, United Kingdom}

\date{Received date; Accepted date}

\abstract{ We present the results of numerical simulations of the
  spherically symmetric gravitational collapse of supermassive stars
  (SMS). The collapse is studied using a general relativistic
  hydrodynamics code.  The coupled system of Einstein and fluid
  equations is solved employing observer time coordinates, by
  foliating the spacetime by means of outgoing null hypersurfaces. The
  code contains an equation of state which includes effects due to
  radiation, electrons and baryons, and detailed microphysics to
  account for electron-positron pairs. In addition energy losses by
  thermal neutrino emission are included.  We are able to follow the
  collapse of SMS from the onset of instability up to the point of
  black hole formation.  Several SMS with masses in the range $5\times
  10^5 M_{\odot}- 10^9 M_{\odot}$ are simulated. In all models an
  apparent horizon forms initially, enclosing the innermost 25\% of
  the stellar mass. From the computed neutrino luminosities, estimates
  of the energy deposition by $\nu\bar{\nu}$-annihilation are
  obtained. Only a small fraction of this energy is deposited near the
  surface of the star, where, as proposed recently by Fuller \& Shi
  (1998), it could cause the ultrarelativistic flow believed to be
  responsible for $\gamma$-ray bursts. Our simulations show that for
  collapsing SMS with masses larger than $5\times 10^5 M_{\odot}$ the
  energy deposition is at least two orders of magnitude too small to
  explain the energetics of observed long-duration bursts at
  cosmological redshifts. In addition, in the absence of rotational
  effects the energy is deposited in a region containing most of the
  stellar mass.  Therefore relativistic ejection of matter is
  impossible.   }

\maketitle

\section{Introduction}

Recent observations with the Space Telescope Imaging Spectrograph on
board the HST have added stronger support to the increasing belief
that supermassive black holes (SMBH) are not unusual in nature: the
central objects in more than 45 galaxies have been examined and in 34
of them the presence of a SMBH has been confirmed (Kormendy 2000; see
also Rees 1998 for an overview of SMBH). The inferred masses of these
objects imply that SMBH of $10^{6}-10^{9}$ $M_{\odot}$ are present in
the center of probably most, if not all, galaxies. Our own Galaxy with
a central mass between $2.6 \times 10^{6} M_{\odot}$ and
$3.3 \times 10^{6} M_{\odot}$ (Genzel {\it et al.} 2000) lies at the
lower limit of this mass range. In the particular cases of NGC 4258
and our Galaxy, the radius inferred for the location of the ``dark
mass'' is sufficiently small to exclude any black hole alternative
(Maoz 1998).  The question of how SMBH form, is, however, still not
settled and the nature of their progenitors is still rather uncertain
(see, e.g., Rees 1984). Two different scenarios have essentially been
proposed, relying upon either stellar dynamics in a dense star cluster
or on the hydrodynamics of a single supermassive star (SMS). Both
cases can develop a dynamical instability (Chandrasekhar 1964; Fowler
1964; Shapiro \& Teukolsky 1985) and can therefore undergo
catastrophic gravitational collapse which would lead to the formation
of a SMBH.

Despite of missing theoretical and observational evidence for the
existence of SMS, it is still of theoretical interest to study their
properties, due to their potential astrophysical relevance as
progenitors of SMBH. In particular, it is worth analyzing the dynamics
involved in the gravitational collapse of such stars. The low
frequency gravitational waves emitted during such (in case of rotating
stars) events could be detected by the proposed Laser Interferometer
Space Antenna (LISA).  Strong bursts of gravitational radiation would
only occur for homologous collapse down to small radii, but not in
case of the formation of an initially small black hole which gradually
grows through subsequent accretion (Thorne \& Braginsky 1976).

A number of collapse simulations of SMS have been attempted through
the years. The first ones were done by Appenzeller \& Fricke (1972a,b)
who studied the collapse of spherical stars of masses $7.5\times
10^{5} M_{\odot}$ and $5.2 \times 10^{5} M_{\odot}$, using a code with
only a subset of the general relativistic equations implemented.
Since core temperatures higher than $5\times10^{8}$ K were
encountered, nuclear burning was expected to release high amounts of
energy. They also found that for stars with masses greater than
$10^{6} M_{\odot}$ thermonuclear reactions have no major effect on the
evolution. In a subsequent paper Fricke (1973) studied the fate of SMS
depending on the initial metallicity and mass. Later, Shapiro \&
Teukolsky (1979) simulated the evolution of a $10^{6} M_{\odot}$ star
with a spherically-symmetric relativistic code which did not include
any microphysics effects. They were able to follow the whole evolution
of the collapsing star until the formation of a black hole.  Fuller,
Woosley \& Weaver (1986) revisited the work of Appenzeller \& Fricke
performing simulations in the range of $10^{5}-10^{6}M_{\odot}$ with a
code which included post-Newtonian corrections and detailed
microphysics.  They found that SMS with zero initial metallicity do
not explode, but the critical metallicity for an explosion was
considerably reduced compared to the results of Appenzeller \& Fricke.
In a series of recent papers, Baumgarte \& Shapiro (1999a,b)
investigated the influence of rotation on the evolution and luminosity
of a SMS, finding that the luminosity is reduced by rotation and,
hence, the lifetime of the star is enhanced.  Furthermore, they showed
that the ratios $R/M,\;T/|W| $ and $ J/M^{2}$, where $R$ is the
radius, $M$ the mass, $T$ the rotational energy, $J$ the angular
momentum and $W$ the gravitational potential energy of the star, are
universal numbers at the onset of instability. Therefore they
concluded that the collapse of SMS should produce universal
gravitational waveforms.

SMS have recently re-gained theoretical interest since Fuller \& Shi
(1998) proposed a scenario to explain the origin of cosmological
$\gamma$-ray bursts (GRBs)
from the collapse of SMS. The large amounts of emitted thermal
neutrinos and antineutrinos during the catastrophic collapse of such
stars (or, alternatively, of stellar clusters) could deposit a
significant fraction of their energy by annihilation to
electron-positron pairs near the surface of the star. They argued that
this annihilation-induced heating could lead to relativistic expansion
and associated $\gamma$-ray emission by cyclotron radiation and/or by
the inverse Compton process, producing the required energies for a GRB
in the range $10^{51}-10^{54}$ ergs within several seconds (assuming
isotropic emission).

In order to give reliable, quantitative estimates of the neutrino
emission from the collapse of SMS we have performed a number of
numerical simulations of spherically symmetric collapsing SMS, using a
relativistic hydrodynamics code (Papadopoulos \& Font 2000a). The code
is able to follow the collapse from the onset of instability up to the
point of black hole formation. This was an important consideration
since, due to the large temperature dependence of the neutrino
emission, most of the neutrinos are emitted close to the onset of
black hole formation, when the temperatures in the stellar core are
highest.

This paper is organized as follows: in Section 2 we present the
problem setup, describing the equation of state we use, the equations
of general relativistic hydrodynamics and the initial data for SMS. We
also discuss some computational issues relevant to our simulations.
These simulations are presented and analyzed in Section 3, together
with some tests to show the accuracy of the numerical code. Moreover,
this section contains our results concerning the neutrino emission and the
energy deposition by neutrino-antineutrino annihilation during 
SMS collapse. In Section 5 we discuss the implications of our
results for GRBs. The paper ends with a summary in Section 6.

\section{Problem setup}

\subsection{Properties of SMS and microphysics}

SMS are supported against gravitational collapse by radiation
pressure, the gas pressure being negligible. Such stars are very well
approximated in equilibrium by polytropes, $p=K\rho^{\Gamma}$,
$\Gamma=1+1/n$ with polytropic index $n=3$ ($\Gamma=4/3$).

SMS have a two-dimensional parameter space: the central density
$\rho_{\mathrm{c}}$, which controls the radius of the star, and the
mass of the star $M$, which determines the ratio $\tilde\beta$ between
gas pressure and total pressure via Eddington's quartic equation
(Kippenhahn \& Wigert 1994)
\begin{equation}
  \label{eq:MassParam}
3.02\times10^{-3}  \left(\frac{M}{M_{\odot}}\right)^2 =\frac{1-\tilde\beta}
{\mu^4\tilde\beta^4},
\end{equation}
where $\mu$ is the mean molecular weight.
The polytropic constant $K$ is related to $\tilde\beta$ by
\begin{equation}
  \label{eq:PolytrConst}
  K=\left( \frac{3\mathcal{R}^{4}}{a\mu^{4}}\right)^{\frac{1}{3}}
\left(\frac{1-\tilde\beta}{\tilde\beta^{4}}\right)^{\frac{1}{3}},
\end{equation}
where $\mathcal{R}$ is the universal gas constant
and $a$ is the radiation constant.

The adiabatic index is slightly larger than $4/3$ due to the
contribution from baryons. The onset of instability occurs when the
adiabatic index drops below a critical value
\begin{equation}
  \label{eq:InstabCritGamma}
  \Gamma'_{\mathrm{crit}}=\frac{4}{3}+1.12 \frac{2 G M}{R c^{2}},
\end{equation}
where $G$ is the gravitational constant and $c$ is the speed of light.
This corresponds to a critical central density (Chandrasekhar 1964)
\begin{eqnarray}
   \label{eq:InstabCritDensity}
  \rho_{\mathrm{crit}}&=&1.994\times10^{18}\left(\frac{0.5}{\mu}\right)^{3}
  \left(\frac{M_{\odot}}{M} \right)^{\frac{7}{2}}\mathrm{g\, cm}^{-3}.
\end{eqnarray}
Stars with a central density $\rho_{\mathrm{c}}$ larger than
$\rho_{\mathrm{crit}}$ will become dynamically unstable.

In our code we have implemented a tabulated equation of state (EOS)
$p=p(\rho,\epsilon)$ which includes radiation, electrons and effects
associated with the creation of electron-positron pairs.
Contributions due to the presence of an additional Boltzmann gas
(H$^{+}$) are taken into account in an appropriate way. Given
$\epsilon$, the specific internal energy, and $\rho$, it is possible
to determine the temperature $T$ by a Newton-Raphson iteration and
then compute the pressure from $\rho$ and $T$.

Furthermore, the EOS allows us to compute neutrino
energy loss rates due to pair annihilation, $e^{-}+e^{+} \rightarrow
\nu + \overline{\nu}$, photo-neutrino emission, $\gamma + e^{\pm}
\rightarrow e^{\pm} + \nu + \overline{\nu}$, and plasmon decay $\gamma
\rightarrow \nu + \overline{\nu}$. The fitting formulas and tables for
the neutrino energy loss rates of the first two processes have been
taken from Itoh {\it et al.} (1996).  The corresponding expressions
for plasmon decay were obtained from Haft {\it et al.} (1994).
These processes become relevant in the very last epochs of the
collapse.

\subsection{General relativistic hydrodynamics}

\begin{figure}[t]
  \begin{center}
    \resizebox{5cm}{!}{\includegraphics{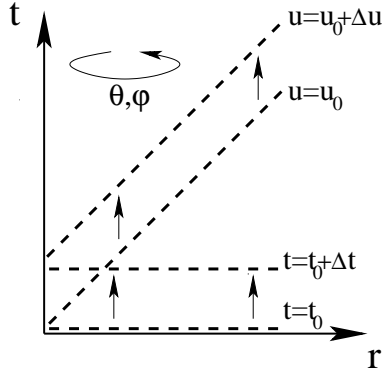}}
    \caption{In the 3+1 formulation initial data are constructed on spacelike
      hypersurfaces, $t\equiv \mathrm{const.}$ In the
      characteristic formulation employed in this work, initial data
      are constructed on null hypersurfaces (lightcones), 
      $u=\mathrm{const}$.}
    \label{fig:lightcone}
  \end{center}
\end{figure}

The formulation and implementation of the Einstein and hydrodynamic
equations follows the work of Papadopoulos and Font (2000a,b). These
equations are formulated adopting an outgoing null foliation 
of the spacetime (Bondi {\it et al.} 1962; Sachs
1962), whose line element, in spherical symmetry reads:
\begin{equation}
  \label{eq:BondiSachs}
  \mathrm{d}s^{2}= - \frac{e^{2 \beta}V}{r}\mathrm{d}u^{2}-2 
  e^{2 \beta}\mathrm{d}u \mathrm{d}r +
  r^{2}(\mathrm{d}\theta^{2} +\sin^{2}{\theta}\mathrm{d}\phi^{2}).
\end{equation}

The radial coordinate $r$ is chosen to make the spheres of rotational
symmetry have an area of $4\pi r^{2}$. The time coordinate represented
by $u$ will be referred to as the ``observer time coordinate''. This
retarded time $u$ labels the null cones in terms of proper time along
the central geodesic.  Additionally, $\theta$ and $\phi$ are angular
coordinates for the null rays, as shown in Figure \ref{fig:lightcone}.
The geometry is completely described by the two metric functions
$V(u,r)$ and $\beta(u,r)$.

The hydrodynamic equations are cast into a first-order, flux-conservative
system which, in spherical symmetry, reads:
\begin{eqnarray}
\partial_u{\bf U} + \frac{1}{\sqrt{-g}}\partial_r{\sqrt{-g}\bf F} = {\bf S},
\end{eqnarray}
with 
\begin{eqnarray}
{\bf U}&=&(D, S^r, E), 
\\
{\bf F}^r&=&(\rho u^r, \rho h u^ru^r +
p g^{rr}, \rho h u^0 u^r + pg^{0r},
\\ 
{\bf S}&=&(-\frac{D}{\sqrt{-g}}\partial_u\sqrt{-g},
-\Gamma^{r}_{\mu\lambda}T^{\mu\lambda} -
\frac{S^r}{\sqrt{-g}}\partial_u\sqrt{-g} ,
\nonumber
\\
& &-\Gamma^{u}_{\mu\lambda}T^{\mu\lambda} -
\frac{E}{\sqrt{-g}} \partial_u\sqrt{-g}), 
\end{eqnarray}
where $T^{\mu\nu}$ is the stress-energy tensor,
$\Gamma^{\nu}_{\mu\lambda}$ are the Christoffel symbols and
$\sqrt{-g}$ is the volume element, which is given, for the chosen
geometry, by $\sqrt{-g}=r^2e^{2\beta}\sin^{2}\theta$.  The state
vector -- the conserved quantities -- contains the relativistic
densities of mass, momentum and energy, defined, in terms of the {\it
  primitive} variables, ${\bf w}=(\rho, u^r, \epsilon)$, as $D=\rho
u^0$, $S^r=\rho h u^0u^r + pg^{0r}$ and $E=\rho hu^0u^0 + pg^{00}$,
respectively. In this expressions $h$ is the specific enthalpy,
$h=1+\epsilon+p/\rho$, and $u^0$ and $u^r$ are the time and
radial components of the 4-velocity, respectively.

The Einstein equations reduce to two radial hypersurface
equations for $\beta$ and $V$:
\begin{eqnarray}
  \label{eq:MetricODEConsbeta}
  \beta_{,r}&=&2\pi r e^{4 \beta} E, \\
  \label{eq:MetricODEConsV}
  V_{,r}&=& e^{2 \beta} - 8 \pi r^{2} e^{4\beta}S^{r} - 4 \pi r
  e^{4\beta}V E.
\end{eqnarray}
The integration of the metric is thus completely separated from the
recovery of the primitive quantities, which simplifies the recovery.
Instead of a three-dimensional root finding procedure, coupled to the
two ODEs for the metric, only a one dimensional problem must be solved.

\subsection{Initial data}

All SMS models in our sample are chosen such that the initial central
density is a bit larger than the critical central density given by
Eq. (\ref{eq:InstabCritDensity}), for the particular mass of the star.
They are listed in Table \ref{tab:params}.  The initial models for the
simulations are therefore described by one parameter only, the mass of
the star.

\begin{table}[t]
\centering
\begin{minipage}{42mm}
\caption{Parameters of the initial models: Mass and central
  density. The criterion for the onset of dynamical instability, Eq.
  (\ref{eq:InstabCritDensity}), gives the central densities.}
\label{tab:params}
\begin{tabular}{*{2}{c}}
\hline
      Mass            \hspace{0.5cm}  & $\rho_{c}$           \\[0.5ex]
      [$M_{\odot}$]   \hspace{0.5cm}  & [g/cm$^{3}$]         \\[0.5ex]
      \hline
      $5\times10^{5}$ \hspace{0.5cm}  & 2.26$\,\times\,10^{-2}$  \\
      $1\times10^{6}$ \hspace{0.5cm}  & 2.80$\,\times\,10^{-3}$  \\
      $5\times10^{6}$ \hspace{0.5cm}  & 7.14$\,\times\,10^{-6}$  \\
      $1\times10^{7}$ \hspace{0.5cm}  & 6.30$\,\times\,10^{-7}$  \\
      $5\times10^{7}$ \hspace{0.5cm}  & 2.26$\,\times\,10^{-9}$  \\
      $1\times10^{8}$ \hspace{0.5cm}  & 1.99$\,\times\,10^{-10}$ \\
      $1\times10^{9}$ \hspace{0.5cm}  & 6.30$\,\times\,10^{-14}$ \\
      \hline
    \end{tabular}
  \end{minipage}
\end{table}

The equations describing the hydrostatic equilibrium are the 
Tolman-Oppenheimer-Volkoff (TOV) equations. Using outgoing Bondi-Sachs 
coordinates they read (Papadopoulos \& Font 2000a):
\begin{eqnarray}
  \label{eq:TOV}
  &p_{,r}&=\left(\frac{1}{2 r} -
   \frac{1}{2 Y}(1+ 8 \pi r^{2} p) \right)\rho h,\\
  &Y_{,r}&=1+8 \pi r^{2}(p-\rho h),
\end{eqnarray}
where $Y=V e^{-2 \beta}$. These coupled differential equations are
integrated by a fourth-order Runge-Kutta scheme. The relation between
pressure and density is prescribed by the polytropic structure
relation. The integration of Eq.  (\ref{eq:MetricODEConsbeta}) yields
$\beta$. The boundary conditions at $r=0$ are
$p=p_{\mathrm{c}}=K\rho_{\mathrm{c}}^{4/3}, Y=0$ and $\beta=0$.  The
polytropic constant can be computed using Equation
(\ref{eq:PolytrConst}) and the corresponding $\tilde\beta$ is given by
Equation (\ref{eq:MassParam}).

It may be criticized that initial data for an unstable star be prescribed 
on a lightcone rather than on a spacelike hypersurface. For stable stars 
the density profile is independent of the spacetime slicing but in dynamic 
situations this does not hold anymore. Other authors (Baumgarte {\it et al.} 
1995) circumvent this problem. They generate initial data on null hypersurfaces
by using a spacelike 3+1 code. They trace outgoing lightrays, and store the
hydrodynamic quantities encountered along its path. However, as the
stellar structure does not change much during the first light-crossing
time, a direct prescription on null hypersurfaces seems to be
sufficient and justified.

\subsection{Computational issues}

The numerical code uses a conservative Godunov-type scheme with an
approximate Riemann solver to integrate the (hyperbolic) hydrodynamic
equations. The hypersurface equations (Eqs.~(\ref{eq:MetricODEConsbeta})
and (\ref{eq:MetricODEConsV})) for the metric components are
solved using a two-step Runge-Kutta scheme. Specific details about the
numerical algorithms can be found in Papadopoulos \& Font (2000a).  The
canonical Eulerian grid used for our collapse simulations of SMS, 
consists of 500 fixed zones distributed in a non-equidistant, geometrically 
growing way from the origin, such that the final Schwarzschild radius of 
the star is resolved with roughly 50 zones.

It is worth discussing in some detail the computational difficulties 
arising from the use of outgoing lightlike spacetime foliations when a black 
hole forms. We have found that in numerical simulations of collapsing SMS 
(see next Section) a black hole forms at the center which only encompasses the 
innermost $\sim$25\% of the total mass. The rest of the star's mass 
accretes on this seed black hole in a dynamical timescale. When the computation 
is continued beyond the moment when the black hole first appears, the density 
profile still looks reasonable, but the internal energy develops spikes and 
grows rapidly in the region $M(r)\ge0.25M$. As expected, in observer time 
coordinates the region interior to that mass does not evolve 
further and thus remains intact. The unphysical behavior can be postponed by
severely limiting the time step near black hole formation. Similar 
observations were reported by G\'omez {\it et al.} (1996) who simulated the
evolution of a scalar field in observer time coordinates up to the point of 
black hole formation. 

\begin{figure}[t]
 \begin{center}
    \resizebox{7.5cm}{!}{\includegraphics{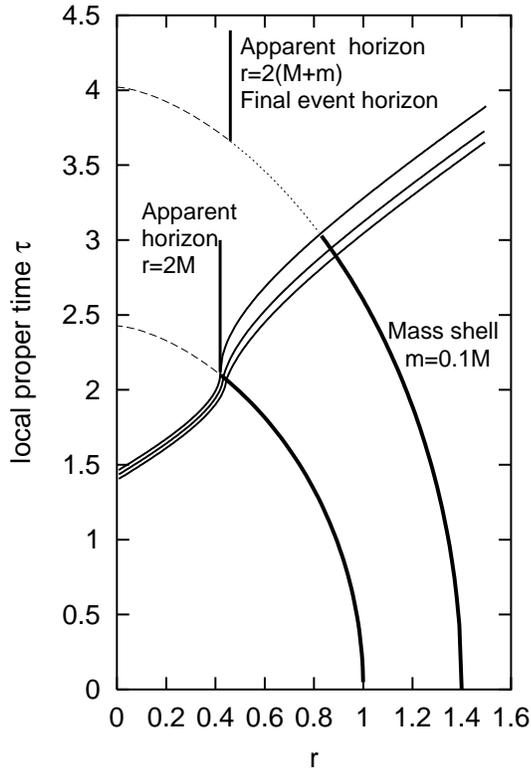}}
    \caption{Spacetime diagram of the
      collapse of a homogeneous dust sphere in arbitrary units. 
      The vertical axis shows the advance of local proper time $\tau(r)$,
      i.e., the time measured by observers at fixed radius $r$.
      The trajectories of three outgoing lightrays (emitted at
      $\tau=1.3978,1.4278,1.4578$) are shown as they escape from the
      forming black hole. Due to the gravitational distortion of the
      spacetime, lightrays are severely delayed close to the apparent
      horizon. Subsequently, a shell of mass $m$ is accreting onto the
      seed black hole.  In such a situation most evolution codes are
      likely to experience problems at the innermost apparent horizon
      due to finite numerical precision. Without taking additional
      measures it is not possible to compute the dotted section of the
      accreting mass shell.  The dashed sections of the spacetime
      diagram are already inside the event horizon and are causally
      disconnected from distant observers. }
     \label{fig:divergingLR}
\end{center}
 \end{figure}

The conditions arising at the onset of black hole formation are shown in 
Figure \ref{fig:divergingLR} for the collapse of a pressureless (dust) star of
homogeneous density.  Lightrays, whose trajectories can be computed analytically,
are emitted from the center of the star. They cross the surface of the
collapsing star when the formation of an apparent horizon is about to
begin and thus they suffer a severe delay.  

The mass $m$ of the accreting mass shell is only important for determining 
the final event horizon, its trajectory being independent of this mass. To 
approach the innermost apparent horizon is most critical since the ``distance'' 
between the apparent horizon and the {\it computational} null hypersurface is 
smallest there. Our numerical code advances data on outgoing null hypersurfaces. 
In Figure \ref{fig:divergingLR}, the three trajectories of the lightrays 
represent the last three time steps of the simulation.  The region below 
the uppermost lightray trajectory, particularly the segments of the stellar 
surface and of the mass shell, drawn by solid lines, have already been 
computed. Without taking additional measures it is not possible to compute the
dotted section of the accreting mass shell.

Other authors have reported similar problems with their implementation of 
the observer time method in the context of the collapse of neutron stars 
to black holes (Baumgarte {\it et al.} 1995). In their case it helped to 
reduce the accuracy of their numerical method by performing a first-order 
accurate integration of the metric component $g_{00}$. Effectively, this 
first-order scheme constantly underestimates $g_{00}$, thus preventing the 
penetration of the horizon. In our formulation of the hydrodynamic equations 
it turns out that manipulating the metric while keeping the conserved 
quantities fixed influences the distribution of the primitive quantities 
in such a way that the metric component $g_{00}$ is less underestimated or 
even overestimated.

Our result that an apparent horizon forms inside the collapsing SMS
at about $M(r)\sim0.25M_{\mathrm{{\small SMS}}}$ agrees quite well with 
the results of other authors. Woosley, Wilson and Mayle (1986) found a 
trapped surface at $M(r)\sim0.2M_{\mathrm{{\small SMS}}}$ for a 
$5\times10^{5}M_{\odot}$ SMS and similar
results were reported by Shapiro \& Teukolsky (1979) for a $10^{6}M_{\odot}$ 
SMS. The slightly smaller value of the mass inside the apparent horizon 
most likely arises from the use of different microphysics.

Our main aim is to compute the lightcurves of neutrino emission during the 
collapse of a SMS. The neutrino luminosities will, like lightrays, suffer 
from severe redshift when black hole formation is in progress. However,
as we discuss later, almost all neutrinos are emitted near the 
center of the star. Therefore, the dominant part of the redshift for
these neutrinos arises from the innermost apparent horizon - the
matter further outside has not yet collapsed far enough, i.e., the outermost
apparent horizon is too distant to have a significant impact on the
redshift.

\section{Simulations}

We turn next to describe our simulations, presenting first a series
of tests we have performed in order to calibrate the code.

\subsection{Tests of the numerical code}

For testing the accuracy of the code we use spherical neutron star 
configurations as there exist either previous results or analytic 
estimates that we can use. Convergence tests show that the code is 
second order accurate in space and time except at the center of the 
star. As the center is a local extremum, the numerical schemes we 
use (which belong to the so-called total variation diminishing class) 
are only first order accurate there.
The code can keep a polytropic model of a neutron star in equilibrium
for very long-term evolutions (roughly 200 light-crossing times).

\begin{figure}[t]
  \resizebox{8cm}{8cm}{\includegraphics{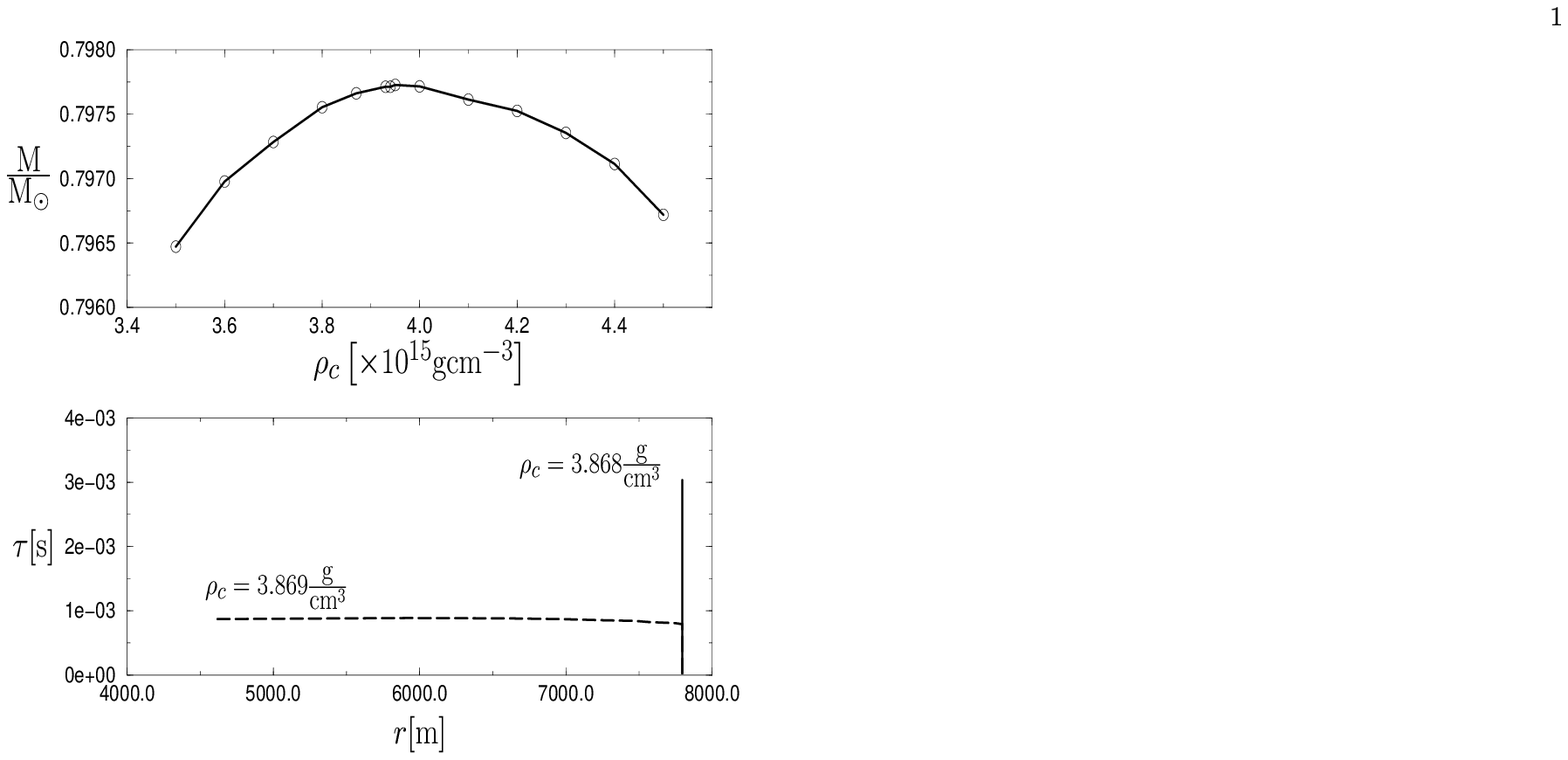}}
   \caption{Test case for determining the onset of instability in a neutron
    star. The upper panel shows the mass against the central density 
    $\rho_{c}$. Models with $\rho_{c}>3.96\times10^{15}$ g cm$^{-3}$ should 
    be dynamically unstable. In the lower panel the surface radii is plotted as
    a function of time for two models with only slightly different central 
    density. While the model with $\rho_{c}=3.868\times10^{15}$ g cm$^{-3}$ 
    is stable and its radius is kept constant in time (solid line), the one with 
    $\rho_{c}=3.869\times10^{15}$ g cm$^{-3}$ collapses to a black hole in less 
    than 1 ms.}
   \label{fig:BaumgMassCurve}
\end{figure}

Determining the onset of instability of stellar models by probing
different central densities provides a very strong test for numerical
codes. In order to do this a sequence of neutron star equilibrium
configurations is constructed by solving the TOV equations for several 
central densities. Only those configurations where $\frac{d M}{d \rho_{c}}>0$ 
are dynamically stable. In order to benchmark our results we use the same 
model as Baumgarte et al (1995), namely a $5/3$ polytrope with
$K=5.38\times10^{9}$ (in cgs units) with a mass of 0.79 $M_{\odot}$.
In Figure \ref{fig:BaumgMassCurve} the dependence of $M$ on $\rho_{c}$
is shown (upper panel). The maximum mass is obtained at 
$\rho_{c}=3.96\times10^{15}$g cm$^{-3}$ .
The lower panel in Figure \ref{fig:BaumgMassCurve} shows the surface radii 
of two different models having almost the
same central density. The model with $\rho_{c}=3.868\times 10^{15}$g
cm$^{-3}$ remains stable (solid line), while the model with 
$\rho_{c}=3.869\times 10^{15}$g cm$^{-3}$ collapses (dashed line).  
This is exactly the same numerical value quoted by Baumgarte et al (1995).
Although the determination of the critical density is only accurate at
2.3$\%$, the corresponding mass is identical up to the first three
significant figures.

The accuracy of the code has also been measured by computing the
frequencies of the radial modes of pulsations of a spherical neutron
star.  The numerical evolution of an initially static star is
influenced by the local truncation error of the hydrodynamic scheme.
These errors excite low amplitude radial pulsations whose frequencies
can be measured by Fourier transforming the time evolution data of any
fluid variable. Since in our code these frequencies are measured in
observer time coordinates (and not in terms of local proper time) one
has to correct them by a redshift factor.  The results can be compared
to computations of linear normal modes (as an eigenvalue problem) or
with results of other nonlinear codes.  Our results agree very well
with the frequencies computed by Font {\it et al.} (2000).  Using 200
equidistant radial zones, the fundamental radial mode and the first two
harmonics are computed with an accuracy of 0.12\%, 3\% and 0.1\%,
respectively.

\subsection{Evolution of collapsing SMS}

We now describe the dynamics of the collapse of a spherical
supermassive star. Although we have computed the collapse of all models
listed in Table \ref{tab:params}, we focus our discussion on the 
$5\times 10^5 M_{\odot}$ star, which represents our canonical model.
Since all SMS are very closely approximated by 4/3-polytropes, the 
evolution of stars of different masses involves different length and 
time scales, but it is otherwise very similar.

\begin{figure*}[t]
\resizebox{18cm}{19cm}{\includegraphics{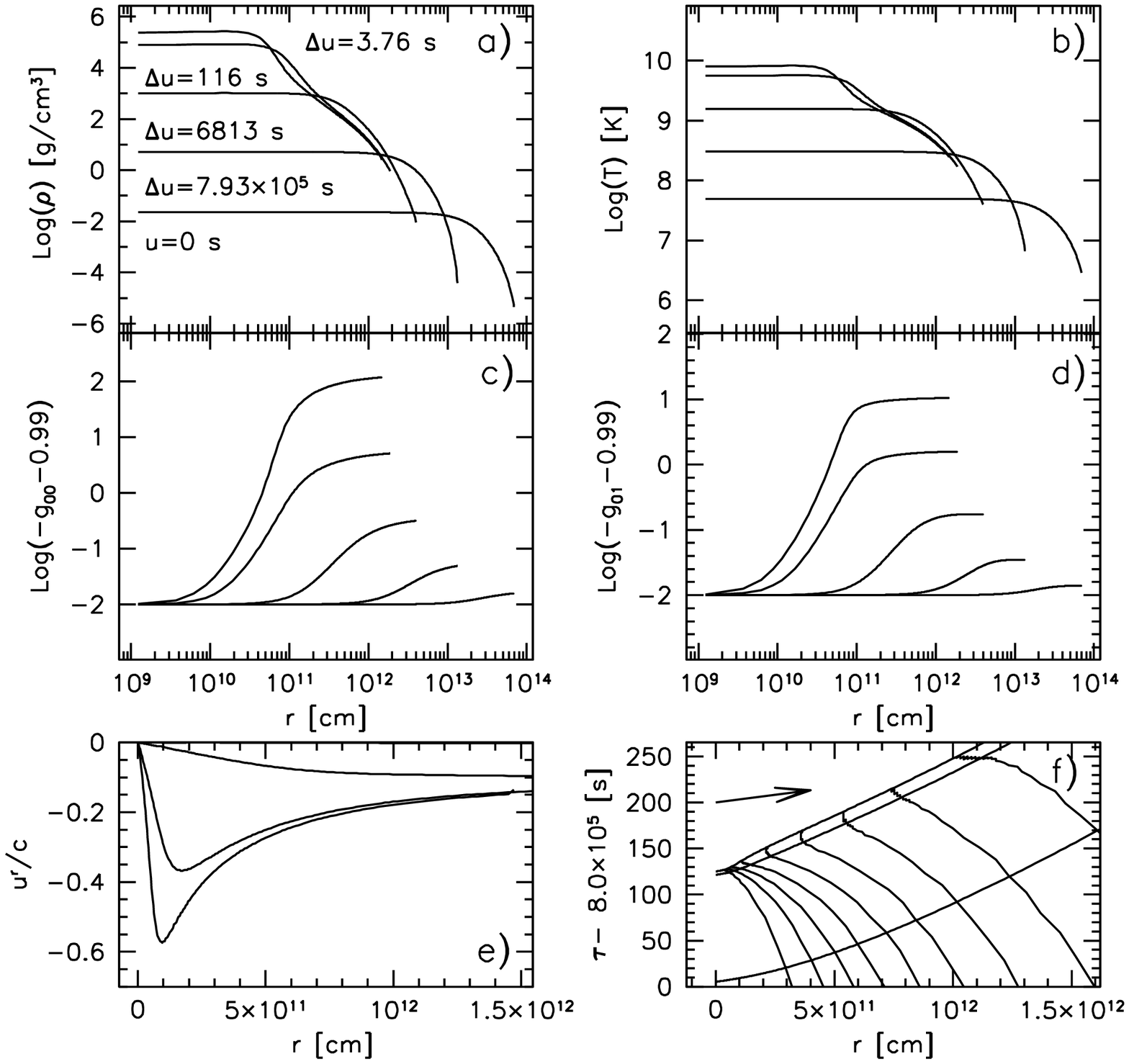}}
      \caption{Snapshots of $\rho$, $T$, $g_{00}$, $g_{01}$ and
        $u^{r}$ are shown for a $5\times10^{5} M_{\odot}$ SMS. The
        coordinate time interval $\Delta u$ elapsed between adjacent
        profiles diminishes since the evolution accelerates. The
        coordinate time is exactly equal to the proper time $\tau$
        elapsed at the center of the star. The deviation of the metric
        from Minkowski spacetime, $g_{0i}=-1$, is shown in the middle
        panels. As it is not possible to display a zero deviation in a
        logarithmic plot, the deviation from $g_{00}=-0.99$ is shown.
        In these plots $\log{(-g_{0\mu}-0.99})\equiv -2$ for
        $\mu\in\{0,1\}$ represents flat spacetime. The plot on the
        lower right side shows the location of mass shells ($\Delta
        M=5\times10^{4}M_{\odot}$) versus local proper time $\tau$. The
        lines intersecting with the mass shells are hypersurfaces of
        constant coordinate time $u$.  They represent trajectories of
        outgoing lightrays. The arrow gives the slope of a lightray at
        $r\rightarrow \infty$.  Gravity already causes a severe delay
        of outgoing lightrays.  Notice the kink in the upper
        hypersurface very close to the center of the star indicating the
        formation of an apparent horizon.  }
      \label{fig:5E5SMS}
\end{figure*}

Figure \ref{fig:5E5SMS} shows the evolution of the density, temperature,
metric components and velocity for the $5\times 10^{5}M_{\odot}$ SMS.  
The collapse lasts $8\times 10^{5}$ s ($\approx 9.3$ days) and the
central density increases by a factor of $1.08\times 10^{7}$.  
The initial configuration can be well described by Newtonian theory, as
$g_{00}(R)=-1.0058$ at the surface of the star $R$, which deviates from the
value in Minkowski spacetime, $g_{00}=-1$, by only 0.58\%. At the end
of the simulation the final configuration
has become highly relativistic and $g_{00}(R)=-119$.  According to
the relation $d\tau^2=(e^{2\beta}V/r) du^2$, this corresponds to a time 
dilation factor of about 11. In order to visualize this dramatic 
increase, Figure \ref{fig:5E5SMS} displays the deviation from flat spacetime
logarithmically. In all simulations the center of the star is chosen
to resemble Minkowski geometry, e.g. the boundary condition for $\beta$
is chosen as $\beta(r\equiv0)=0$. Due to the use of observer time coordinates 
all hydrodynamic quantities seem to freeze at the end of the simulation
while the metric quantities start to grow very rapidly when black hole
formation is about to begin.  The maximum radial velocity encountered
is about $-0.57c$ at the very end of the simulation. The velocity
profile is proportional to $r$ up to $r\approx 5.6\times10^{10}$ cm,
which corresponds to about 25\% of the star's mass.

The panel on the lower right side of Figure \ref{fig:5E5SMS} shows the
local proper time $\tau(r)$ versus the position of mass shells enclosing
fixed fractions of the total mass of the star. Between two lines the
enclosed mass increases by 10\% of the total mass of the star.  The
three lines intersecting the trajectories of the mass shells represent
hypersurfaces of constant coordinate time $u$.  They can be
interpreted as the trajectories of outgoing lightrays. The arrow
indicates the slope of a lightray at asymptotically large radii, i.e.,
in Minkowski spacetime.  Thus, lightrays are severely delayed due to
the distortion of the spacetime.  By comparing the position of the
mass shell enclosing 90\% of the mass of the star with the
Schwarzschild radius of a $5\times10^{5}M_{\odot}$ star,
$R_{s}=1.47\times10^{11}$ cm, it becomes clear that the star still has
to contract to less than one sixth of its current radius.

The density profiles in Figure \ref{fig:5E5SMS} are self-similar,
i.e., the collapse proceeds homologously.  A perfect homologous
collapse is found until the central density has increased by a factor
$798$ corresponding to a central density $\rho_{\mathrm{c}}\approx 18$
g/cm$^{3}$. At $\rho_{\mathrm{c}}\approx10^{3}$ g/cm$^{3}$ the
deviation from homology is evident. Goldreich \& Weber (1980) found
analytically that in Newtonian theory a star with polytropic index
$\Gamma=4/3$ should contract strictly homologously. The deviation from
homologous collapse observed in the simulation is mainly due to
effects of general relativity: time dilation results in a slower
advance of local proper time $\tau(r)$ near the center of the star,
and the proper volume element $\sqrt{-g}dr$ is larger than in
Newtonian gravity.  Therefore the density profiles in Figure
\ref{fig:5E5SMS} tend to become steeper. Another contribution comes
from the use of detailed microphysics.  Due to copious $e^{-}e^{+}$
pair production at the center of the star, where temperatures are
highest, the increase of the internal energy is used to create the
rest mass of the leptons instead of generating additional pressure.
The EOS is thus softened, i.e. the adiabatic index is reduced, and the
star tends to become more centrally condensed. However, since the
deviation from homology is already apparent when the central
temperatures are lower than $10^9$K (see Figure \ref{fig:5E5SMS}),
$e^{-}e^{+}$ pair creation can be considered as a minor contribution.

The evolution of SMS of different masses is found to be qualitatively
very similar. The mass essentially affects the evolution only as a scaling 
parameter, except for the effects of neutrino emission. 
In all models of our sample, a black hole forms from the innermost 
25\% of the total stellar mass, and thus, the location of the innermost
apparent horizon is proportional to $M$. In addition,
the velocity profiles at the onset of black hole formation are
found to be nearly identical. The timescale $\Delta\tau$ of neutrino emission,
$\Delta\tau\approx R_{s}/u^{r}(R_{s})$, is thus proportional to the stellar mass 
$M$ (see also Figure \ref{fig:Luminosities}). The dependence of the final
core temperature on the mass is found to be $T_{c}\sim M^{-0.5}$ 
(Shi \& Fuller 1998).

\begin{figure}[t]
\begin{center}
\resizebox{7cm}{!}{\includegraphics{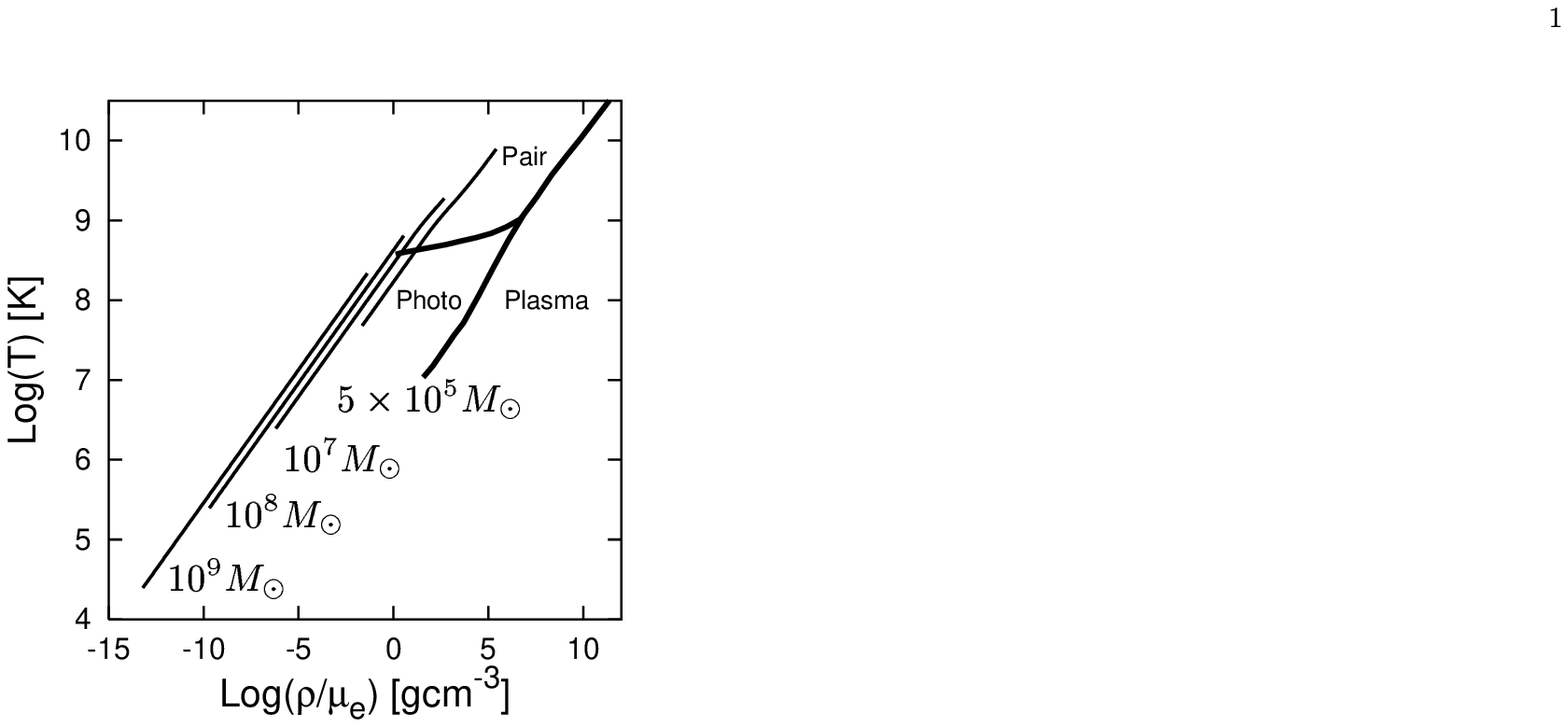}}
  \caption{Evolutionary tracks of several simulated SMS in the
    $\rho-T$-plane. More massive stars develop lower final core
    temperatures at the onset of black hole formation.  }
  \label{fig:NeutrinoProcess2}
\end{center}
\end{figure} 

\subsection{Neutrino emission}

As already mentioned, three different kinds of thermal processes of
neutrino production are included in the simulations: $e^{+}e^{-}$-pair
annihilation, photo-neutrino emission, and plasmon decay. The neutrino
generation rates in thermodynamic equilibrium depend only on the
temperature and density of the matter.  Figure
\ref{fig:NeutrinoProcess2} shows the evolutionary tracks, including
effects due to electron-positron pair creation in the stellar gas at $T>10^{9}$K, for
our sample of stars in the range of
$5\times10^{5}M_{\odot}-10^{9}M_{\odot}$.  

From Figure \ref{fig:NeutrinoProcess2} it is obvious that the plasmon
decay process is negligible for the simulations. For stars with masses
below $5\times10^{8}M_{\odot}$ the pair creation process will be the
most important process, as most of the energy is released in the last
decade of central density increase.  Black hole formation prevents
stars with masses larger than $10^{9}M_{\odot}$ from entering the
region dominated by pair annihilation. Neutrinos are then emitted
primarily by the photo-neutrino process. Since the neutrino emission
rate per volume for pair annihilation is $Q_{\nu}\sim T^9$
$[\mathrm{erg/s/cm}^{3}]$ (Itoh {\it et al.}  1996), one can already
infer from Figure \ref{fig:NeutrinoProcess2} that the total energy
release in form of neutrinos decreases with increasing stellar mass.

\begin{figure}[t]
  \begin{center}
    \resizebox{8.0cm}{!}{\includegraphics{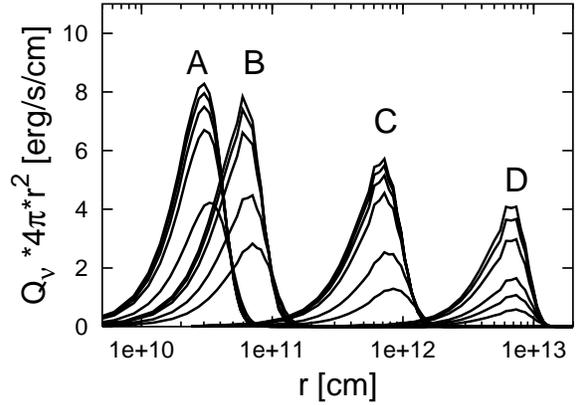}}
    \caption{Radial dependence of the (scaled) differential neutrino luminosities 
      $\mathrm{d}L_{\nu\overline{\nu}}(r)/\mathrm{d}r$ for four SMS:
      (A) $5\times10^{5}M_{\odot}$, (B) $10^{6}M_{\odot}$, (C)
      $10^{7}M_{\odot}$ and (D) $10^{8}M_{\odot}$.  The various lines
      indicate different epochs for each model (increasing upwards).
      The upper curves correspond to the end of the simulations. The
      scale factors used are $5\times10^{44}$ (A), $10^{44}$ (B),
      $2\times10^{41}$ (C) and $2\times10^{36}$ (D), respectively. The
      main fraction of the neutrinos is emitted around a radius
      $R_{\nu}$, which is much smaller than the radius of the star at
      the end of the simulation ($\sim1.47\times10^{12}$cm for the
      $5\times10^{5}M_{\odot}$ SMS).}
    \label{fig:DiffLum}
  \end{center}
\end{figure}

As the volume increases $(\sim r^2\mathrm{d}r)$ and the temperature
decreases with radius, most neutrinos are emitted from a narrow
spherical shell deep inside the star. Figure \ref{fig:DiffLum} depicts
the radial dependence of the differential neutrino luminosities, $\mathrm{d}
L_{\nu\overline{\nu}}(r)/\mathrm{d}r=4\pi r^2 Q_{\nu}$, for four models of our
sample at various epochs (represented by the different lines for each
star). The shells are centered at a radius $R_{\nu}$, at which the
differential luminosity has a maximum. The total (unredshifted)
neutrino luminosity for all flavors
\begin{equation}
L_{\nu\overline{\nu}}=L_{\nu_{e}}+ L_{\overline{\nu}_{e}} + L_{\nu_{\mu}}+
L_{\overline{\nu}_{\mu}}+L_{\nu_{\tau}}+ L_{\overline{\nu}_{\tau}},
\end{equation}
is given by
\begin{equation}
  \label{eq:DefL}
  L_{\nu\overline{\nu}}(t)=\int_{0}^{\infty} 4\pi r^{2}
  Q_{\nu}\left(T(r,t)\right) \mathrm{d}r.
\end{equation}

\begin{figure}[t]
\begin{center}
  \resizebox{7.5cm}{!}{\includegraphics{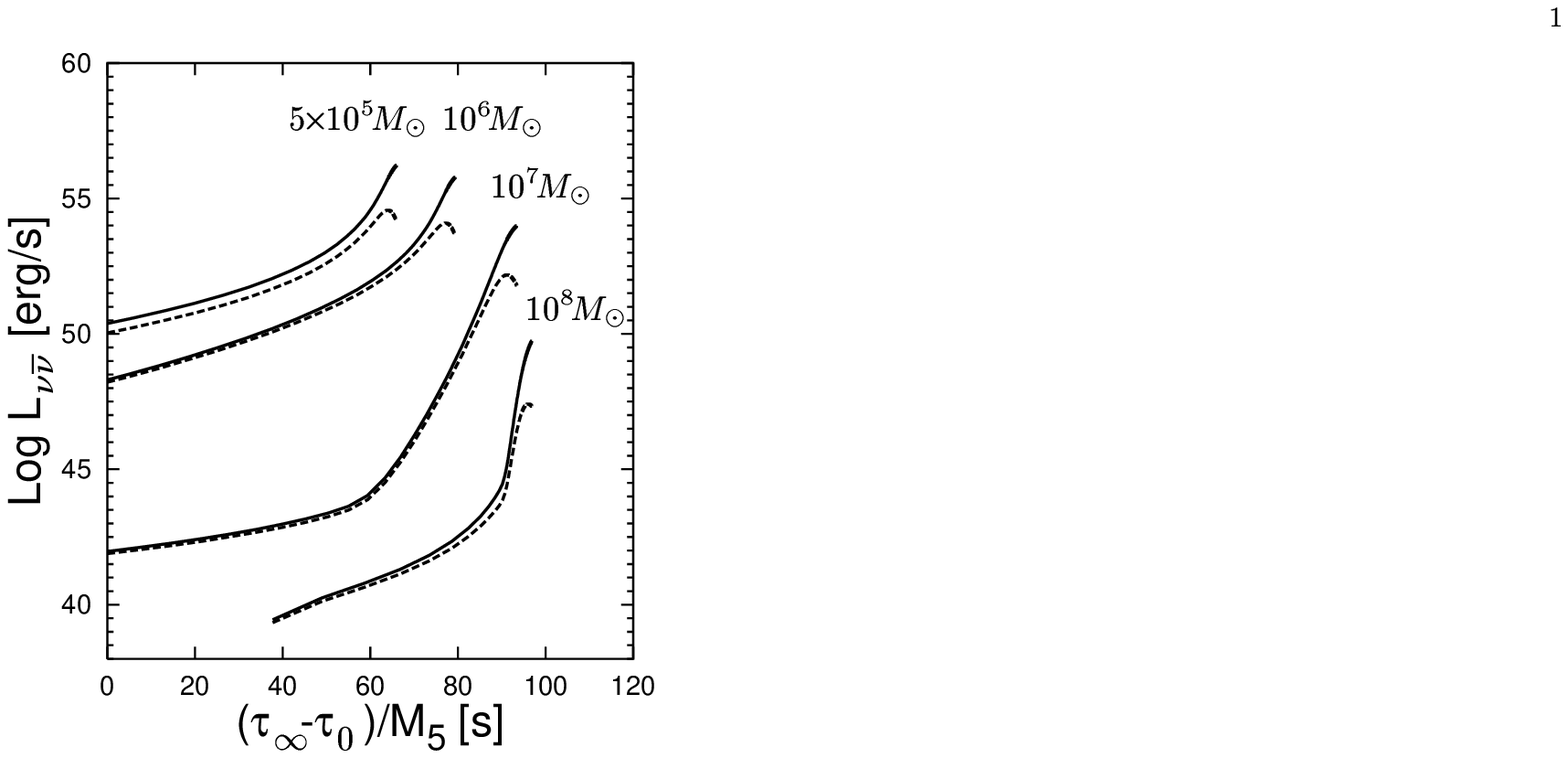}}
  \caption{Time evolution of the neutrino luminosities, bare (solid) and 
    redshifted (dashed), for several stars in the range
    $5\times10^{5}M_{\odot}-10^{8}M_{\odot}$. Quantity $\tau_{\infty}$
    represents the proper time for an observer at infinity. This time
    is measured relative to the overall collapse timescale
    $\tau_{0}=\{8.05\times10^{5}$s,$1.7\times10^{6}$s,
    $8.05\times10^{7}$s,$3.2\times10^{9}$s$\}$
    and scaled by $M_{5}=M/(10^{5}M_{\odot})$. As expected, the
    maximum luminosity decreases with the mass of the star.  The
    change of the slope of the neutrino light curves at luminosities
    $\sim 10^{43}$ erg/s, visible in the models with $10^{7}M_{\odot}$
    and $10^{8}M_{\odot}$, is due to the transition from
    photo-neutrino dominated to pair-dominated emission. The rates of
    energy release by neutrino production is much more sensitive to
    the temperature in the latter case.}
  \label{fig:Luminosities}
\end{center}
\end{figure}

Figure \ref{fig:Luminosities} shows the computed neutrino luminosities
$L_{\nu\overline{\nu}}(\tau_{\infty})$ as seen by an observer at
infinity for the collapse of different SMS.  
For each stellar mass the bare and the redshifted neutrino
luminosities are depicted.  It is possible to identify the moment in
time when gravitational redshift overcompensates the increasing
neutrino energy release rates caused by growing temperatures. At the
end of the simulation the luminosities are redshifted by more than two
orders of magnitude.

\begin{figure}[t]
  \begin{center}
    \resizebox{7.5cm}{!}{\includegraphics{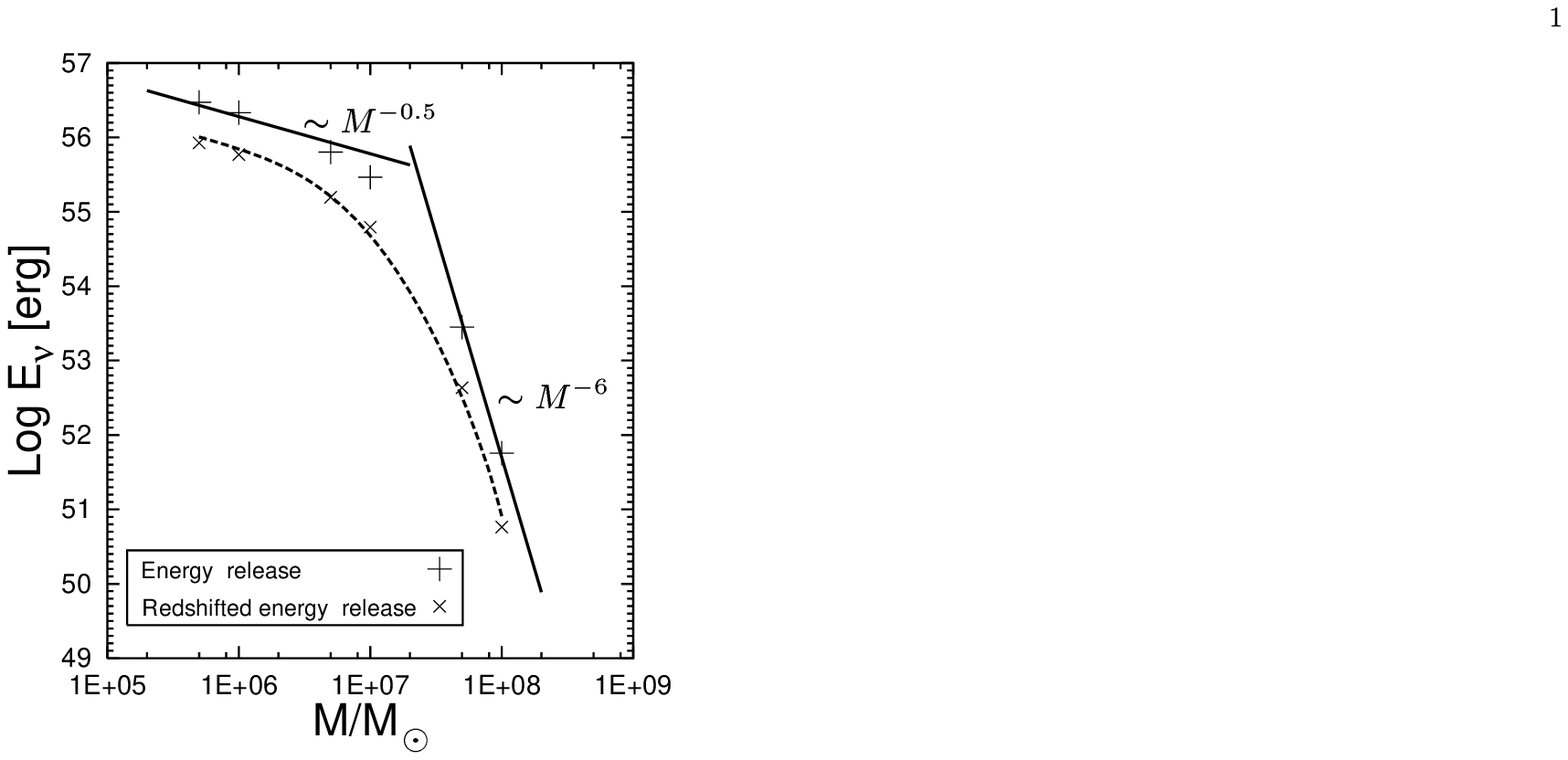}}
    \caption{The total energy release in form of neutrinos against the
      mass of the SMS. The symbols indicate the computed models. The
      data points can be described by a broken power law. }
    \label{fig:MassEtot}
  \end{center}
\end{figure}

\begin{figure}[t]
  \begin{center}
    \resizebox{7.5cm}{!}{\includegraphics{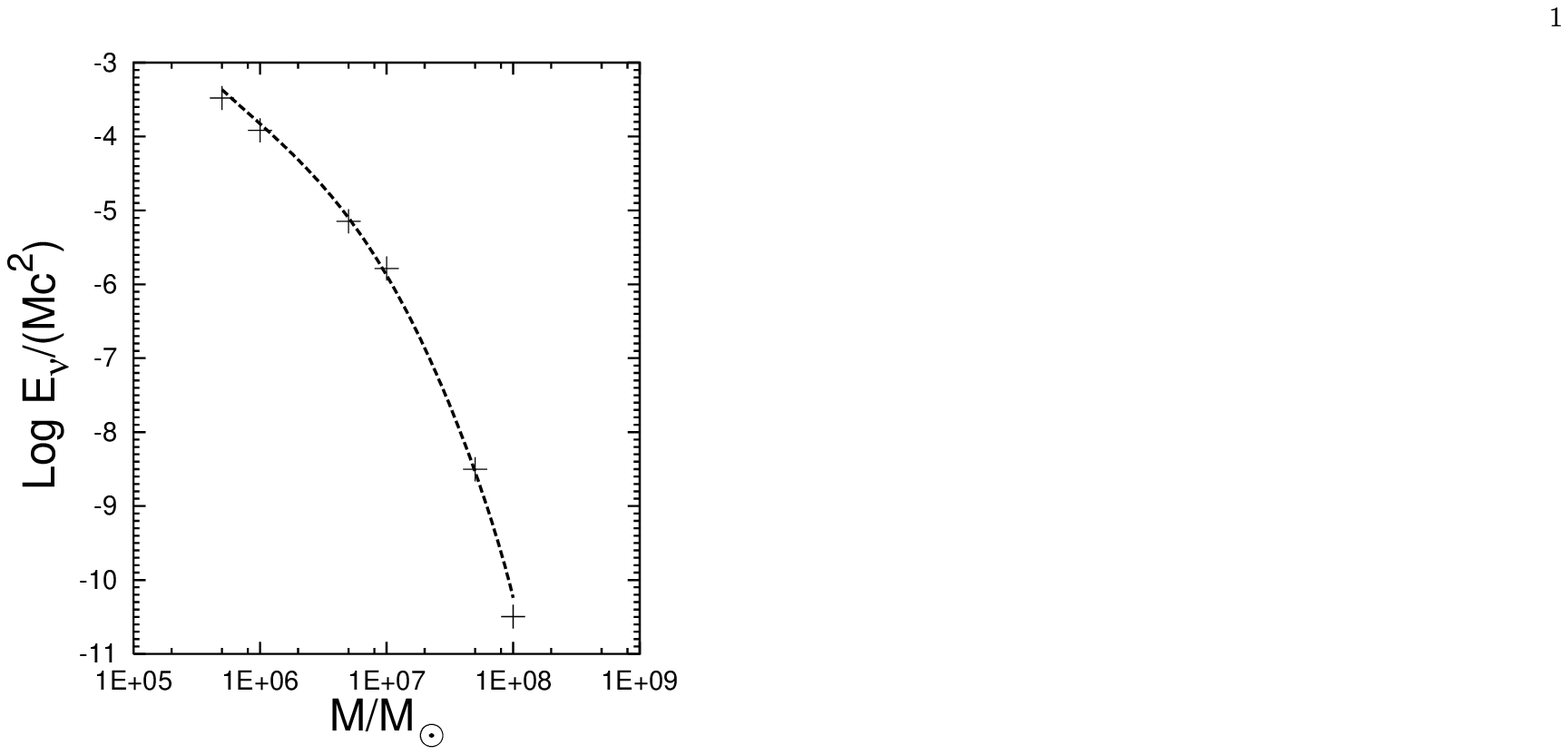}}
    \caption{Efficiency for converting the rest mass of a SMS to neutrinos. 
      As less massive stars evolve to
      higher core temperatures, both the total energy emitted in form
      of neutrinos and the conversion efficiency, rise with decreasing
      mass.}
    \label{fig:MassRM_eff}
  \end{center}
\end{figure}

The total energy release in form of neutrinos during the collapse is
shown in Figure \ref{fig:MassEtot} as a function of the stellar mass.
The data points can be approximated by two different power laws.
Doppler shift and gravitational redshift reduce the plotted values by
approximately a factor of five. We find that for a
$5\times10^{5}M_{\odot}$ SMS the total radiated energy amounts to
$3.0\times10^{56}$ ergs, corresponding to an efficiency for converting
rest mass energy to neutrino energy of $3.4\times10^{-4}$. This
efficiency is depicted in Figure \ref{fig:MassRM_eff} and defined as the
(unredshifted) energy released in neutrinos divided by the total rest-mass
energy of the star. After the formation of the black hole and after we 
stopped our simulations, we do not expect a significant fraction of the 
neutrino energy to be emitted.
The reasons for this expectation will be discussed in some detail 
below. Therefore the efficiency is normalized to the rest-mass energy of
the whole star, $Mc^2$. 

These results are in good agreement with the
simulation of Woosley, Wilson \& Mayle (1986), who included neutrino
transport during the collapse of a $5\times10^{5}M_{\odot}$ SMS with
zero metallicity. They found a total energy output of
$2.6\times10^{56}$ ergs in form of electron antineutrinos.  As the
neutrino energy release in the pair-dominated region depends on the
ninth power of the temperature, most neutrinos are emitted, in the
case of a $5\times10^{5}M_{\odot}$ SMS, from the very center of the
star within the last $\Delta \tau_{\infty}\approx10$ seconds of the
collapse before a black hole forms.  Even slightly different
core temperatures at the onset of black hole formation can result in
a large difference of the total energy release.

The broken power law behavior can be understood in terms of 
simple considerations. Assuming adiabatic contraction, $T_\nu \sim
M^{-0.5}$, a typical timescale of neutrino emission $\Delta\tau\sim M$
(see Figure \ref{fig:Luminosities}), a radiating volume with radius
$R_{\nu}\sim R_{s}\sim M$ (see Figure \ref{fig:DiffLum}) and a
neutrino energy emissivity proportional to the ninth power of the
temperature (pair annihilation) at $R_{\nu}$ (Itoh {\it et al.}
1996), the dependence of the total energy release on the mass of the
star is given by $E_{\nu}\sim M^{-0.5}$ (Fuller \& Shi 1998).  This
dependence describes very well the results of the simulations in the
low mass range (high core temperatures $T_{\nu}>3\times10^{9}$ K).
Stars with masses above $5\times10^{7}M_{\odot}$ reach core
temperatures just below $10^{9}$ K where the neutrino emissivity is
proportional to the 20th power of the temperature because of the onset
of $e^{+}e^{-}$-pair creation (Itoh {\it et al.} 1996). The total
energy release is thus proportional to $E_{\nu}\sim M^{-6}$, which is
again in good agreement with the simulations.  In summary, one finds
for the total neutrino luminosities the limiting relations
\begin{eqnarray}
L_{\nu\overline{\nu}}\sim Q_{\nu}
\frac{4\pi}{3}R_{\nu}^{3}\sim \left\{
 \begin{array}{r@{ \;\; }l}
   M^{-1.5} & (10^{5}\lesssim\frac{M}{M_{\odot}}\lesssim
   5\times10^{6})\,,  \\[0.2cm]
   M^{-7} &(5\times10^{7}\lesssim\frac{M}{M_{\odot}}\lesssim10^{8})
   \,,
\end{array} \right.
\label{neulum}
\end{eqnarray}
and for the total energy release (not redshifted)
\begin{eqnarray}
E_{\nu}\sim \left\{
 \begin{array}{r@{ \;\; }l}
   M^{-0.5} & (10^{5}\lesssim\frac{M}{M_{\odot}}\lesssim
   5\times10^{6}) \,, \\[0.2cm]
   M^{-6} &(5\times10^{7}\lesssim\frac{M}{M_{\odot}}\lesssim10^{8})
   \,.
\end{array} \right.
\label{eulum}
\end{eqnarray}

\subsection{$\nu\overline{\nu}$-annihilation and energy deposition}

In 1998 Fuller \& Shi proposed the collapse of a SMS as a possible
mechanism to trigger GRBs. In order to check
this idea quantitatively, we calculate in this section the amount
of energy that is deposited by $\nu\overline{\nu}$-annihilation in
our collapse models of SMS. In the following, the results obtained for
the neutrino emission in the previous section will also be used to
estimate the spatial distribution of the deposited energy. 

Our hydrodynamic simulations indicate that, provided nuclear burning
is unimportant, the gravitational collapse of a SMS proceeds rather
smoothly, involving no shocks.  The collapse leads to the formation of
a black hole consisting of the innermost 25\% of the star's mass, onto
which the rest of the stellar matter accretes. At the same time an
energy up to $\mathcal{O}(10^{57})$ ergs is released in form of
neutrinos. A small fraction of this energy is deposited within the
star. In order to accelerate stellar matter to ultrarelativistic
speeds, the energy deposition must take place in an almost baryon-free
region, i.e. very close to the surface of the star, or in a baryon-poor
funnel along the polar axis in case of a rotating SMS.  The
kinetic energy of the ultrarelativistic flow could then be converted
to $\gamma$-rays by cyclotron radiation and/or by the inverse Compton
process.

Such ultrarelativistic flows from SMS might be powered by the energy deposition
due to $\nu\overline{\nu}$-annihilation (Fuller \& Shi 1998),
\begin{equation}
  \nu_{e,(\mu,\tau)} + \overline{\nu}_{e,(\mu,\tau)}\rightarrow e^{-} + e^{+},
\end{equation}
since this process is capable to deposit a large amount of energy
even in vacuum.  The duration of the strong neutrino emission from the
collapse of a $5\times10^5M_\odot$ SMS is about 30 seconds for an
observer at infinity (see Figure \ref{fig:Luminosities}) and thus in
agreement with the duration of observed long GRBs (Piran 1999).

For a neutrino-radiating sphere of radius $R_{\nu}$, the 
volume-integrated energy deposition rate due to
$\nu\overline{\nu}$-annihilation above $R_{\nu}$ is (Goodman {\it
  et al.} 1987; Cooperstein {\it et al.} 1987)
\begin{equation}
  \dot{E}_{\mathrm{dep}}^{e,(\mu,\tau)}= \frac{A}{R_{\nu}}
  L_{\nu_{e(\mu,\tau)}}
  L_{\overline{\nu}_{e(\mu,\tau)}}
  \left(\frac{\langle\epsilon_{\overline{\nu}}^{2}\rangle}
  {\langle\epsilon_{\overline{\nu}}\rangle }+ 
  \frac{\langle\epsilon_{\nu}^{2}\rangle}{\langle\epsilon_{\nu}\rangle } 
  \right),
\end{equation}
with $A=K_{e,(\mu,\tau)} G_{F}^{2}/9\pi c$. Here
$G_{F}^{2}=2.066\times10^{-32}$ cm$^{2}$/erg$^{2}$ is the Fermi
constant, $L_{\nu_{e(\mu,\tau)}}$ and
$L_{\overline{\nu}_{e(\mu,\tau)}}$ are the neutrino and antineutrino
luminosities at $R_{\nu}$, $\langle\epsilon_{\nu}\rangle$ and
$\langle\epsilon_{\nu}^{2}\rangle$ energy moments of the neutrino
phase space distribution, evaluated by using a fitting formula for the
pair neutrino spectrum as given by Shi \& Fuller (1998).
$K_{e,(\mu,\tau)}$ is a dimensionless constant, which, according to the
standard model of weak interactions, is given by
\[ K_{e,(\mu,\tau)}=\left\{ 
  \begin{array}{r@{\; \mathrm{for} \;}l} 
    \frac{1}{6\pi} \left(1+4\sin^2{\theta_{W}}+8\sin^4{\theta_{W}}\right) &
    \nu_{e}\overline{\nu}_{e},\\[0.2cm]
    \frac{1}{6\pi}\left(1-4\sin^2{\theta_{W}}+8\sin^4{\theta_{W}}\right) &
    \nu_{\mu,\tau}\overline{\nu}_{\mu,\tau},
  \end{array} 
\right. \] with $\theta_{W}$ being the Weinberg angle
($\sin{\theta_{W}}=0.23$). Assuming that most neutrinos are produced
by $e^{-}e^{+}$-pair annihilation it is possible to compute the total
energy deposition rate $\dot{E}_{\mathrm{dep}}^{e(\mu,\tau)}$ as
\begin{eqnarray}
  \label{eq:nu_eval}
  \dot{E}_{\mathrm{dep}}^{e,(\mu,\tau)}= 2A
    \left(\frac{K_{e,(\mu,\tau)}}{K_{e}+2 K_{(\mu,\tau)}}
    \frac{L_{\nu\overline{\nu}}}{2} \right)^{2}
  \frac{41.225}{5.769}\frac{T_{\nu}}{R_{\nu}},
\end{eqnarray}
where $T_{\nu}$ is the temperature at $R_{\nu}$ and
$L_{\nu\overline{\nu}}$ is the total neutrino luminosity.

For the sake of simplicity and comparison with Fuller \& Shi (1998),
all effects due to relativistic redshift or general relativistic ray
bending were neglected in the evaluation of the
$\nu\overline{\nu}$-annihilation. Towards the end of our simulations,
however, they play an increasingly important role. For the
$5\times10^{5}M_{\odot}$ star, the main neutrino emitting region is
located near $3\times10^{10}$cm. According to Figure
\ref{fig:5E5SMS}c, the gradient of the redshift factor is steepest 
around that radius,
close to which the apparent horizon forms. This gradient will become
even steeper and grow even faster as the collapse proceeds. Gravitational
redshift will therefore quench the neutrino luminosity, also in the 
layer where most of the neutrinos annihilate. At larger radii the
temperature is too low to allow for strong neutrino production. This
will in particular be true during the phase when the forming black hole swallows
the remaining $\sim 75$\% of the SMS, because there is no resistance to the
infalling gas, and compressional heating will therefore stay moderate. For this 
reason it is very unlikely that the accretion of the rest of the star's mass
onto the seed black hole will yield a large contribution to the total
neutrino emission and energy release by $\nu\bar\nu$-annihilation.

\begin{figure}[t]
  \begin{center}
    \resizebox{7.5cm}{!}{\includegraphics{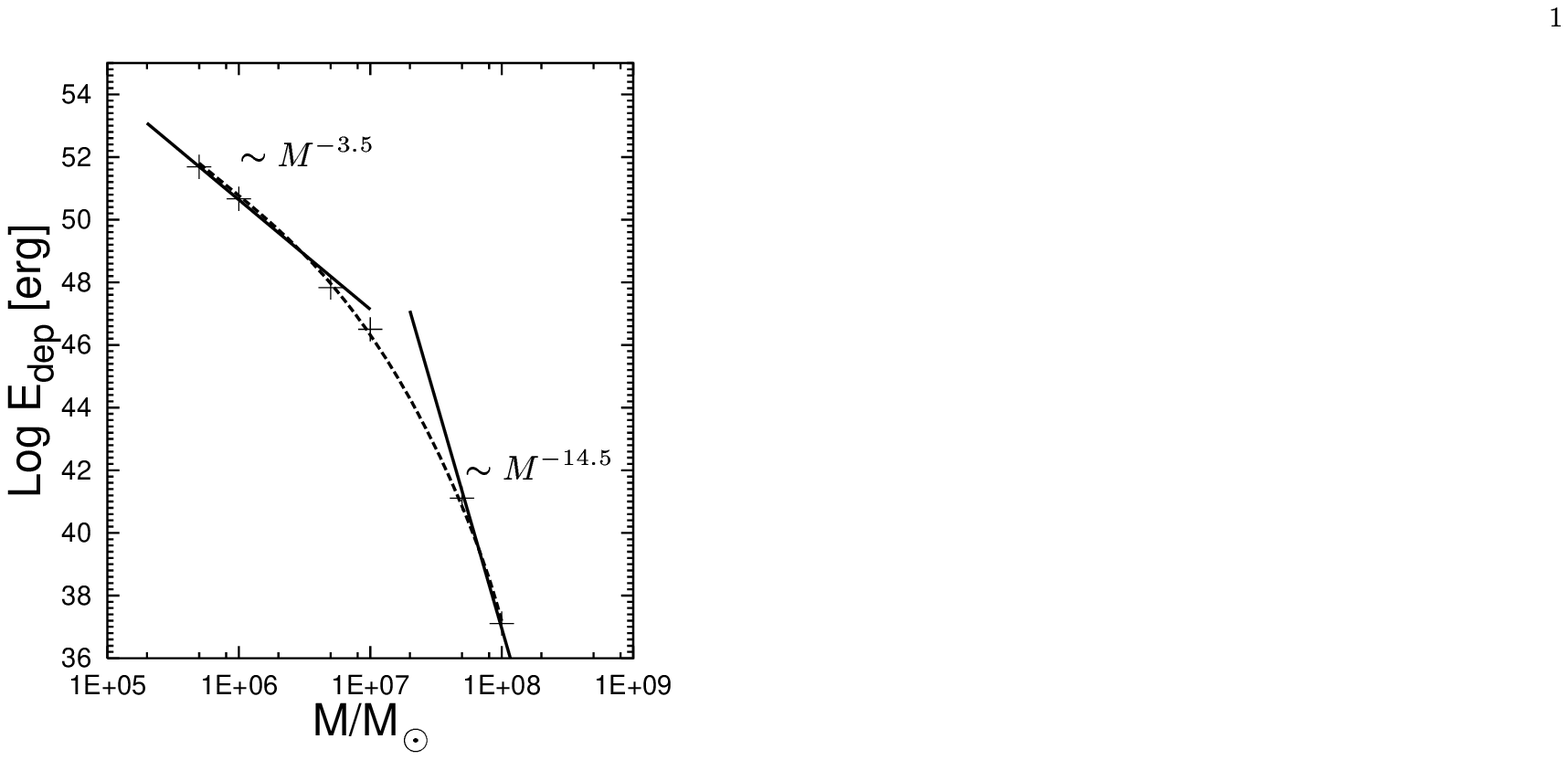}}
    \caption{The total neutrino energy deposition as a function of the mass of
      the SMS.  Due to decreasing neutrino luminosities the total
      energy deposited within the star decreases rapidly with the mass
      of the star. The dependences on both extremes of the considered
      mass range can be understood by simple considerations (see text
      for details). }
    \label{fig:MassEnergyDepos}
  \end{center}
\end{figure}

\begin{figure}[t]
  \begin{center}
    \resizebox{7.5cm}{!}{\includegraphics{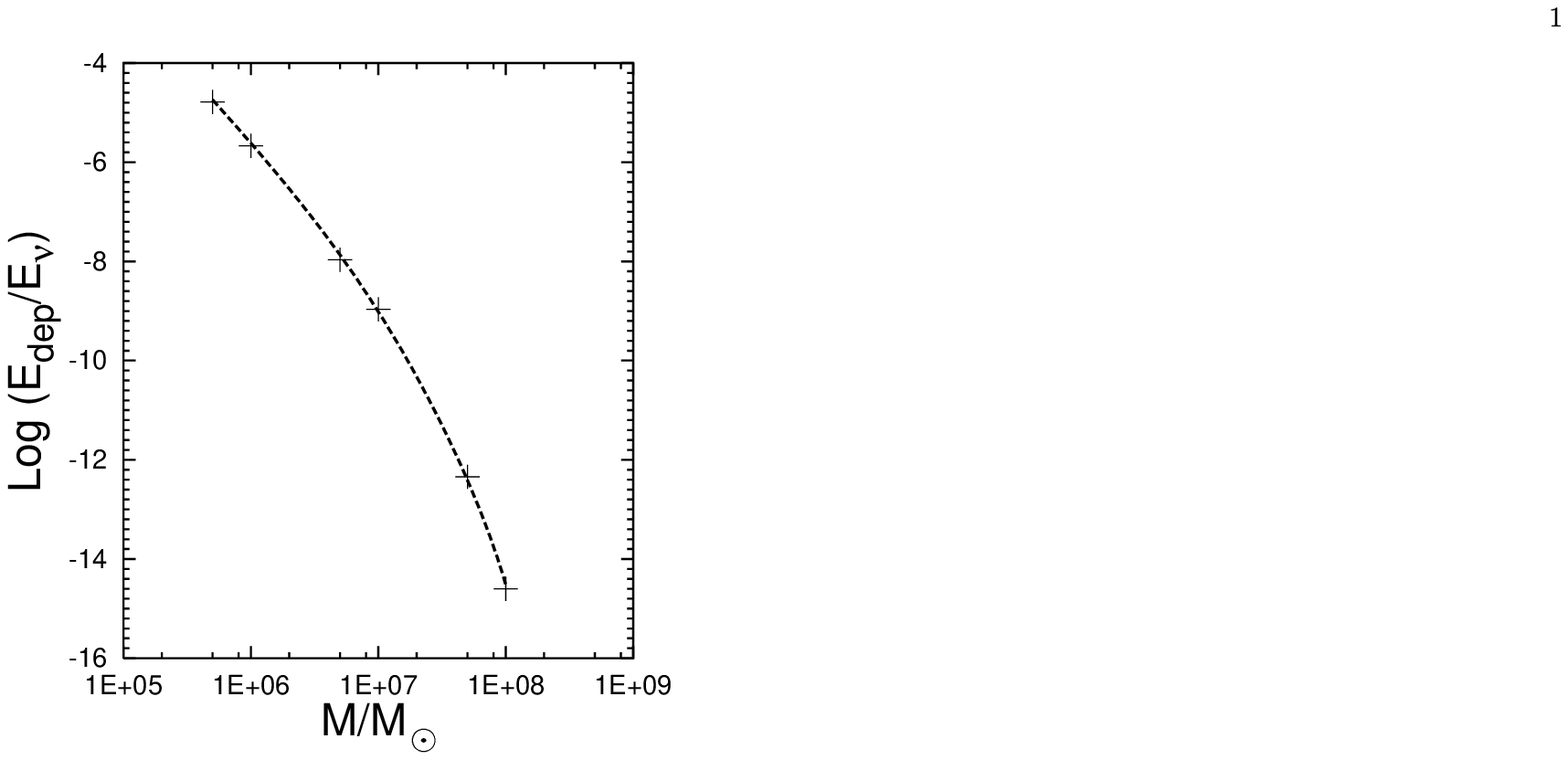}}
    \caption{Efficiency for conversion of neutrino energy to pair plasma by
      $\nu\overline{\nu}$-annihilation.}
    \label{fig:Mass_Edep_eff}
  \end{center}
\end{figure}

The resulting energy deposition for different SMS is
plotted in Figure \ref{fig:MassEnergyDepos}. The corresponding energy
deposition efficiency $E_{\mathrm{dep}}/E_\nu$ is shown in Figure
\ref{fig:Mass_Edep_eff}. The dependence of the deposited energy on the
mass of the star at both ends of the considered mass range is 
$E_{\mathrm{dep}}\sim L_{\nu\overline{\nu}}^{2} T_{\nu}
\Delta\tau/R_{\nu} \sim L_{\nu\overline{\nu}}^{2}M^{-0.5}$, where
$\Delta\tau$ is the timescale of the emission. Therefore, using
Eq.~(\ref{neulum})
\begin{eqnarray}
  E_{\mathrm{dep}}
  \sim \left\{ \begin{array}{r@{\;\quad}l} M^{-3.5} &
            (10^{5}\lesssim\frac{M}{M_{\odot}}\lesssim
      5\times10^{6})\,, \\ [0.2cm]
      M^{-14.5} & (5\times10^{7}\lesssim\frac{M}{M_{\odot}}\lesssim10^{8})
      \,.
  \end{array} \right.
\end{eqnarray}

Fuller \& Shi (1998) gave analytic estimates for the expected
neutrino emission from the collapse of a SMS. They employed a simple
model to deduce the final average core temperature $T_{\mathrm{c}}$ as a function
of the mass of the homologously collapsing core $M^{\mathrm{HC}}$. Assuming
neutrino energy
release rates $Q_{\nu} \sim T_{\mathrm{c}}^9$ and a radiating sphere with a radius
equal to the Schwarzschild radius, $R_{\mathrm{s}}=2G M^{\mathrm{HC}}/c^2$, for the
homologously collapsing mass, their estimate for the characteristic
neutrino luminosity is
\begin{equation}
  \label{eq:FullerL}
  L_{\nu\overline{\nu}}\approx 5\times10^{57} (M_5^{\mathrm{HC}})^{-3/2}\;\;
  \mathrm{erg/s},
\end{equation}
where $M_5^{\mathrm{HC}}=M^{\mathrm{HC}}/10^5M_{\odot}$.
Furthermore, assuming a typical emission timescale of
$\Delta\tau\approx M_5^{\mathrm{HC}}$ s, their result for the energy deposited above a
radius $r$ is
\begin{equation}
  E_{\mathrm{dep}}\approx2.5\times10^{54} (M_5^{\mathrm{HC}})^{-3.5} (R_{\mathrm{s}}/r)^5
  \;\;\mathrm{erg}.
\end{equation}

The simulation presented in the previous section shows that in a
$5\times10^5M_\odot$ SMS an inner core with 25\% of the stellar mass
is collapsing homologously, i.e. $M_5^{\mathrm{HC}}=1.25$. Table
\ref{tab:FullerComp} shows the estimates of Fuller \& Shi in comparison 
to the results of our simulation.  Since the physics
involved in the description of SMS is well known, all their basic
assumptions have been confirmed by our simulation: (1) The core
temperature $T_{\mathrm{c}}\sim T_\nu\sim M^{-0.5}$, (2) the radius of the
radiating sphere $R_\nu\sim M$, (3) the time scale of the emission
$\Delta\tau \sim M$, and (4) the computed neutrino energy
generation rates do not differ by more than 25\% if they are evaluated
for the same temperature $T\approx10^{10}$ K.  The large discrepancy
of the total deposited energy $E_{\mathrm{dep}}$ is caused by the
strong dependence of the energy deposition rate
$\dot{E}_{\mathrm{dep}}$ on $T_{\nu}$ and $R_{\nu}$,
$\dot{E}_{\mathrm{dep}}\sim T_\nu^{19} R_{\nu}^{5}$ (see Eq.
\ref{eq:nu_eval} with Eq.~\ref{neulum}). 
Misjudging slightly both the radius $R_{\nu}$ of
the radiating sphere and the temperature $T_\nu$ at that radius leads
to a very large overestimation of the energy deposition rate. Furthermore,
the way in which we compute the neutrino luminosities in our
simulations, Eq.~(\ref{eq:DefL}), is also more accurate than the
formula employed by Fuller \& Shi, $L_{\nu\overline{\nu}}=
Q_{\nu}\left(T_{\nu}\right)\frac{4\pi}{3}R_{\nu}^{3}$. This, however,
decreases the ratio of Fuller and Shi's neutrino luminosity to our 
luminosity by a factor of three, corresponding to a reduction of the
disagreement of the energy deposition by $\nu\bar\nu$-annihilation by
a factor of nine, to yield a remaining discrepancy of nearly a factor of 600.

\begin{table}[t]
\centering
\begin{minipage}{78mm}
  \caption{Comparison of the analytic model of Fuller \& Shi (1998)
   with the results of our numerical simulation. Values are missing
   where the approximations by Fuller \& Shi are not applicable or
   do not provide information.  Overestimating the temperature
   $T_\nu$ at the neutrino-radiating sphere $R_\nu$ leads to a large
   error in the rate of energy release by neutrino emission $Q_\nu$
   and thus in the energy deposition $E_{\mathrm{dep}}$ at $3
   R_{\nu}\le r < \infty$.  } 
 \label{tab:FullerComp}
 \begin{tabular}{*{3}{c}} \hline $M=5\times10^5M_\odot$& Simulation
   & Fuller \& Shi \\ 
   \hline 
   $T_\nu$ [K] & $8.0\times10^9$ & $11.6\times10^9$ \\ 
   $R_\nu$ [cm] & $3.0\times10^{10}$ & $3.8\times10^{10}$ \\ 
   $Q_\nu(R_\nu)$ [erg/s/cm$^3$] & $4.1\times10^{23}$ & $1.2\times10^{25}$ \\   
   $L_{\nu\overline{\nu}}$ [erg/s] & $1.7\times10^{56}$& $3.6\times10^{57}$ \\
   $E_{\mathrm{dep}}(r>R_\nu)$ [erg] &  $4.9\times10^{51}$& --- \\ 
   $E_{\mathrm{dep}}(r>3 R_\nu)$ [erg] & $8.0\times10^{48}$& $4.7\times10^{51}$\\
   $M(r>3 R_\nu)$ & $\sim 0.65 M_{\mathrm{SMS}}$ & --- \\ 
   \hline 
 \end{tabular}
\end{minipage}
\end{table}

\section{Implication for gamma ray bursts}

The integral amount of deposited energy,
$E_{\mathrm{dep}}\approx4.9\times10^{51}$ erg in the case of a
$5\times10^{5}M_{\odot}$ SMS, is at the lower end of the observed
energy released in cosmic GRBs. All of this energy would have to be
collimated into narrow jets with a collimation of
$\delta\Omega/4\pi\lesssim 1/1000$---$1/100$ in order to explain an equivalent
isotropic energy of $10^{53}$ to $\ga 10^{54}$ erg as inferred for observed
long-duration bursts with known cosmological redshifts (e.g., Frail 
{\it et al.} 2001). However, 99.837\% (Goodman
{\it et al.} 1987) of this energy is deposited in a spherical layer
deep inside the star at $R_{\nu}\le r \le 3 R_\nu$, and only a tiny
fraction near the surface of the star, where excessive baryon loading
could be avoided.  Therefore ultrarelativistic ejection of matter with
Lorentz factors $\gamma\gg 1$ cannot be expected in spherical models:
\begin{equation}
  \gamma\approx1+\frac{E_{\mathrm{dep}}(r>R_\nu)}{M(r>R_\nu)
  c^{2}}\approx1.
\end{equation}
We therefore conclude that energy deposition by
$\nu\overline{\nu}$-annihilation in the spherical collapse of a
$5\times10^{5}M_{\odot}$ SMS does \emph{not} meet the demands for
being a successful central engine for a GRB.

A rotating SMS would have a reduced baryon density along the
polar axis and the time scale of neutrino emission could be longer,
because a disk stabilized by centrifugal forces would prolong the
accretion of matter into a forming black hole. Thus, the conditions 
to fulfill the constraints of GRB models might appear more suitable in this
respect. Nevertheless, as the energy release rates by neutrino
emission and the energy deposition in non-rotating models are several
orders of magnitude below the estimates of Fuller \& Shi, it still
seems unlikely that the total energy release in a baryon-poor
environment would be sufficiently large in rotating scenarios.
However, an ultimate statement requires detailed numerical simulations
of the gravitational collapse of rotating SMS.

As shown in Figure \ref{fig:MassEnergyDepos} the total energy
deposition depends strongly on the mass $E_{\mathrm{dep}}\sim
M^{-3.5}$. Therefore, more massive stars are less promising to meet
the energy requirements of a GRB. Extrapolating the total deposited
energy of a $5\times10^{5}M_{\odot}$ star to less massive stars, the
expected value for a $10^{5}M_{\odot}$ SMS is
$E_{\mathrm{dep}}\approx1.3\times10^{54}$ erg.  The energy deposited
within $[3R_{\nu},\infty]$ is a factor $1.63\times10^{-3}$ smaller and
therefore of the order of $10^{51}$ ergs, which is at the lower end of
the observed energy range. However the baryon loading problem is still
critical. We have not attempted to simulate the evolution of such a
star because in this case nuclear burning becomes relevant.  The
energy release in the form of neutrinos will also depend strongly on
the fate of the star: If the energy release due to nuclear burning
cannot inhibit black hole formation the neutrino emission increases
with smaller stellar mass. If nuclear energy is liberated rapidly enough,
the star will be destroyed in a thermonuclear explosion. Consequently, there
will be much less neutrino emission, since the core temperature will
be considerably lower.  Only detailed numerical simulations of the
collapse of such stars can determine the expected energy release and
their final fate.

\section{Summary}

We have presented results of numerical simulations of the spherically
symmetric gravitational collapse of supermassive stars. Such
simulations were performed using a general relativistic hydrodynamics
code.  The coupled system of Einstein and fluid equations was solved
adopting a spacetime foliation with outgoing null (characteristic)
hypersurfaces.  The code includes a tabulated equation of state which
accounts for contributions from radiation, electrons, electron-positron pairs and
baryonic gases.  Energy losses by thermal neutrino emission were taken
into account.

We were able to follow the collapse of SMS from the onset of
instability up to the point of black hole formation.  Several SMS with
masses in the range of $5\times 10^5 M_{\odot}- 10^9 M_{\odot}$ were
simulated, showing that an apparent horizon forms in all cases,
enclosing the innermost 25\% of the stellar mass. This is in good
agreement with previous simulations by other authors. We did not
attempt to follow the subsequent accretion of the remaining mass onto
the central black hole because, due to finite numerical precision, the
innermost apparent horizon is ultimately penetrated (see Figures
\ref{fig:divergingLR} and \ref{fig:5E5SMS}, in particular the little
kink developing in the lightcones in frame (f) of the latter). This
effect however, does not interfere with the main aim of our work,
namely the computation of neutrino luminosities, because most
neutrinos are emitted near the very center of the star. Therefore we
did not try possible remedies to follow the evolution for longer
times, such as excising the inner regions of the domain, in which the
evolution essentially freezes, and imposing appropriate boundary
conditions there.

The neutrino luminosities for several SMS models have been computed
and it has been possible to understand the dependence of the total
energy release in neutrinos, $E_{\nu}$, on the stellar mass $M$ in terms
of simple scaling laws. Based on these results we have given estimates
of the energy deposition by $\nu\bar{\nu}$-annihilation in the star.
Our simulations show that for collapsing SMS with masses larger than
$5\times 10^5 M_{\odot}$ this energy deposition is more than two
orders of magnitude smaller than the estimates of Fuller \& Shi
(1998). In addition, all of this energy is deposited deep inside the
star and cannot drive relativistic outflows.  Therefore, the spherical
collapse of such a SMS does not meet the demands for being a
successful GRB model.
 
\begin{acknowledgements}
  All computations were performed on the NEC SX-5/3C supercomputer at
  the Rechenzentrum Garching.  We want to thank Maurizio Salaris who
  kindly provided us with his implementation of the energy
  production rates by neutrino processes as published by Itoh {\it et
    al.} (1996) and Haft {\it et al.} (1994). This work was supported
  in part by the SFB-375 ``Astroparticle Physics" of the Deutsche
  Forschungsgemeinschaft.
\end{acknowledgements}

{}


\begin{thebibliography}{}

\bibitem[1972a]{af1972a}
Appenzeller, I., Fricke, K.,
1972, A\&A, 18, 10

\bibitem[1972b]{af1972b}
Appenzeller, I., Fricke, K.,
1972, A\&A, 21, 285

\bibitem[1995]{bst1995}
Baumgarte, T.W., Shapiro, S.L., Teukolsky, S.A.,
1995, ApJ, 443, 717

\bibitem[1999a]{bs1999a}
Baumgarte, T.W., Shapiro, S.L.,
1999, ApJ, 526, 937

\bibitem[1999b]{bs1999b}
Baumgarte, T.W., Shapiro, S.L.,
1999, ApJ, 526, 941

\bibitem[1962]{b1962}
Bondi, H., van der Burg, M.J.G., Metzner, A.W.K.,
1962, Proc. R. Soc. London, Sect. A 269, 21

\bibitem[1964]{chandra1964}
Chandrasekhar, S.,
1964, ApJ, 140, 417

\bibitem[1987]{c1987}
Cooperstein, J., van den Horn, L.J., Baron, E.,
1987, ApJ, 321, L129

\bibitem[2000]{fsk2000}
Font, J.A., Stergioulas, N., Kokkotas, K.D.,
2000, MNRAS, 313, 678

\bibitem[1965]{fowler1964}
Fowler, W.F.,
1964, Rev. Mod. Phys., 36, 545 

\bibitem[2001]{frail2001}
Frail, D.A., {\em et al.}, 2001, preprint, astro-ph/0102282

\bibitem[1973]{fricke1973}
Fricke, K.,
1973, ApJ, 183, 941

\bibitem[1998]{fs1998}
Fuller, G.M., Shi, X.,
1998, ApJ, 502, L5

\bibitem[1986]{fww1986}
Fuller, G.M., Woosley, S.E., Weaver, T.A.,
1986, ApJ, 307, 675 

\bibitem[2000]{gpego2000}
Genzel, R., Picho, C., Eckart, A., Gerhard, O.E., Ott, T.,
2000, MNRAS, 317, 348

\bibitem[1980]{gw1980}
Goldreich, P., Weber, S.V.,
1980, ApJ, 238, 991

\bibitem[1987]{gdn1987}
Goodman, J., Dar, A., Nussinov, S.,
1987, ApJ, 314, L7

\bibitem[1996]{glpw1996}
G\'omez, R., Laguna, P., Papadopoulos, P., Winicour, J.,
1996, Phys. Rev. D, 54, 4719

\bibitem[1994]{hrw1984}
{Haft}, M., {Raffelt}, G., {Weiss}, A.
1994, ApJ, 425, 222

\bibitem[1996]{itoh1996}
{Itoh}, N., {Hayashi}, H., {Nishikawa}, A., {Kohyama}, Y.,
1996, ApJ Suppl., 102, 411

\bibitem[1994]{k1994}
Kippenhahn,R. and Wigert,A.,
1994, Stellar Structur and Evolution, Springer

\bibitem[2000]{k2000}
Kormendy, J.,
2000, preprint, astro-ph/0007401

\bibitem[1998]{maoz1998}
Maoz, E.,
1998, ApJ, 494, L181

\bibitem[2000a]{pf2000a}
Papadopoulos, P., Font, J.A., 
2000a, Phys. Rev. D, 61, 024015

\bibitem[2000b]{pf2000b}
Papadopoulos, P., Font, J.A., 
2000b, preprint, gr-qc/9912054

\bibitem[1999]{pt1999}
Piran, T., 
1999, Phys. Rep., 314, 575

\bibitem[1984]{r1984}
Rees, M.J.,
1984, Ann. Rev. Astr. Astrphys., 22, 471

\bibitem[1998]{r1998}
Rees, M.J.,
1998, in Black Holes and Relativistic Stars, ed. R.M. Wald
(University of Chicago Press, Chicago)

\bibitem[1962]{sachs1962}
Sachs, R.K.,
1962, Proc. R. Soc. London, Sect A 270, 103

\bibitem[1979]{st1979}
Shapiro, S.L., Teukolsky, S.A.,
1979, ApJ, 234, L177

\bibitem[1985]{st1985}
Shapiro, S.L., Teukolsky, S.A.,
1985, ApJ, 298, 34

\bibitem[1998]{sf1998}
Shi, X., Fuller, G.\ M.,
1998, ApJ, 503, 307

\bibitem[1976]{tb1976}
Thorne, K.S., Braginsky, V.B.,
1976, ApJ, 204, L1

\end{thebibliography}
\end{document}